# Robust two-dimensional superconductivity and vortex system in $Bi_2Te_3$/FeTe heterostructures

Hong-Chao Liu[1,2], Hui Li[1], Qing Lin He[3], Iam Keong Sou[1,3], Swee K. Goh[2,*], and Jiannong Wang[1,3,*]

[1]*Department of Physics, The Hong Kong University of Science and Technology, Clear Water Bay, Hong Kong, China*

[2]*Department of Physics, The Chinese University of Hong Kong, Shatin, New Territories, Hong Kong, China*

[3]*William Mong Institute of Nano Science and Technology, The Hong Kong University of Science and Technology, Clear Water Bay, Hong Kong, China*

*corresponding author: skgoh@phy.cuhk.edu.hk, phjwang@ust.hk

The discovery of two-dimensional superconductivity in $Bi_2Te_3$/FeTe heterostructure provides a new platform for the search of Majorana fermions in condensed matter systems. Since Majorana fermions are expected to reside at the core of the vortices, a close examination of the vortex dynamics in superconducting interface is of paramount importance. Here, we report the robustness of the interfacial superconductivity and 2D vortex dynamics in four as-grown and aged $Bi_2Te_3$/FeTe heterostructure with different $Bi_2Te_3$ epilayer thickness (3, 5, 7, 14 nm). After two years' air exposure, superconductivity remains robust even when the thickness of $Bi_2Te_3$ epilayer is down to 3 nm. Meanwhile, a new feature at ~13 K is induced in the aged samples, and the high field studies reveal its relevance to superconductivity. The resistance of all as-grown and aged heterostructures, just below the superconducting transition temperature follows the Arrhenius relation, indicating the thermally activated flux flow behavior at the interface of $Bi_2Te_3$ and FeTe. Moreover, the activation energy exhibits a logarithmic dependence on the magnetic field, providing a compelling evidence for the 2D vortex dynamics in this novel system. The weak disorder associated with aging-induced Te vacancies is possibly responsible for these observed phenomena.

**Introduction**

In condensed matter physics, exotic physical phenomena usually emerge at the heterostructure interface of two materials with different topological characters. Recently, new topological materials including topological insulators,[1-13] topological superconductors,[1, 2, 14-17] and topological semimetals,[18-22] have attracted considerable attention owing to the presence of novel physical properties with promising applications in spintronics, quantum computing, valleytronic devices *etc*. Unlike conventional insulators, three-dimensional (3D) topological insulators, *e.g.* $Bi_2Se_3$ and $Bi_2Te_3$, are protected by the time reversal symmetry and have insulating bulk surrounded by metallic surface states with helical Dirac fermions. At the interface between the topological surface states and an *s*-wave superconductor, a two-dimensional (2D) spinless $p_x + ip_y$ topological superconductor was predicted to be induced by the proximity effect, which hosts the Majorana fermions.[14] In order to observe the Majorana fermions in condensed matter systems, the proximity effect has been widely investigated in topological insulator-superconductor devices[23-27] and heterostructures[28-32]. Adopting a different strategy, we prepared and reported the 2D interfacial superconductivity in topological insulator-iron chalcogenide, *i.e.*, $Bi_2Te_3$/FeTe heterostructure, where neither $Bi_2Te_3$ nor FeTe thin films was superconducting.[33] Since superconductivity can be stabilized in a heterostructure containing the topological insulator $Bi_2Te_3$, their interplay is expected to provide a new platform for the search of Majorana fermions, which obey the non-Abelian statistics and may play an important role in the development of fault-tolerant quantum computer.[34] Furthermore, in topological superconductors the Majorana bound states are always hosted in the vortex cores.[14] Therefore, the stability of the superconductivity and vortex dynamics study become important topics for the search and further manipulation of the Majorana fermions.

Here, we study the robustness of the interfacial superconductivity and 2D vortex dynamics in the $Bi_2Te_3$/FeTe heterostructure using the electrical transport measurements. After two years' air exposure, the aged heterostructures remain superconducting but with a broader transition region. A new feature appears at ~13 K after two years, and its relevance to superconductivity is further revealed by high field studies. The resistance of the $Bi_2Te_3$/FeTe heterostructure just below the transition temperature follows the Arrhenius relation, which we attribute to thermally activated flux flow (TAFF) behavior. The activation energy exhibits a logarithmic dependence on the applied magnetic field, indicating the existence of a 2D vortex system.

**Results and Discussion**

Figure 1(a) presents the normalized temperature dependent resistances $R(T)$ of $Bi_2Te_3$(7 nm)/FeTe in both as-grown and after-two-years measurements. Before the normalization, the normal state resistance of all after-two-years samples is generally higher: using $R$(18 K) as a benchmark, it is 12% - 64% larger compared with the as-grown samples. Since the formation of Te vacancies is always observed in the $Bi_2Te_3$ and FeTe after long-term exposure to air,[35-37] it provides an explanation for the aging induced resistance increase here. Despite the increase in resistance, superconductivity is still robust after two years' air exposure, although the transition regime becomes broader and the zero-resistance temperature $T_{zero}$ drops. Therefore, the weak disorder from the Te vacancies is not completely detrimental to superconductivity; instead it provides an interesting avenue for investigating the vortex dynamics, which will be discussed later. Compared with the as-grown result, in $Bi_2Te_3$(7 nm)/FeTe the temperature of the maximum resistance, $T_{max}$, shifts from 12.4 K to 13 K and the resistance exhibits a two-step drop below $T_{max}$ in the after-two-years case. To provide a clearer view on these results, $dR/dT$ curves are plotted in Figure 1(b). As can be seen, in the after-two-years case, a shoulder appears at around 11 - 13 K, corresponding to the first slow resistance drop in the $R(T)$ curve. As temperature further decreases, a sharp transition starts. We define the starting point of the sharp resistance drop as $T_{mid}$ in after-two-years $R(T)$ result, as shown in Figure 1(b). Therefore, compared with the as-grown result, it demonstrates that an additional new feature appears at around 13 K in after-two-years case.

To learn more about the observation of the new feature, we further measure $R(T)$ of after-two-years heterostructures with different $Bi_2Te_3$ layer thicknesses. Compared with the as-grown results in Fig. 1(c) and its inset, a broader superconducting transition and a two-step resistance drop are observed in all after-two-years heterostructures even with the $Bi_2Te_3$ layer down to 3 nm as shown in Fig. 1(d) and its inset. Furthermore, from the $dR/dT$ curves in the inset of Fig. 1(d), $T_{max}$ shows an increase after two years and locates at ~ 13 K for all samples, manifesting that the new feature is indeed induced by the aging effect. Meanwhile, relative to the full superconducting transition, the shoulder is weakened as the $Bi_2Te_3$ thickness decreases as shown in the inset of Fig. 1(c) and (d), which indicates that the new feature around 13 K is probably relevant to the interfacial superconductivity of the heterostructure.

To further investigate the aging effect on the superconducting transition of $Bi_2Te_3$/FeTe heterostructure, the temperature dependent resistances in different magnetic fields are studied. Figures 2(a) and (b) show the as-grown and after-two-years $R(T)$ results of $Bi_2Te_3$(14 nm)/FeTe in different magnetic fields applied perpendicular to the *ab* plane, respectively. With an increasing magnetic field, $T_{max}$, $T_{mid}$ and $T_{zero}$ gradually shift to lower temperature together as superconductivity is suppressed. Similar behavior is also observed in the parallel fields. From the $R(T)$ curves in different magnetic fields, the corresponding upper critical fields $H_{max}$, $H_{mid}$ and $H_{zero}$ in both as-grown and after-two-years cases can be obtained, and the $H$ - $T$ phase diagrams of Bi$_2$Te$_3$(7 nm)/FeTe and Bi$_2$Te$_3$(14 nm)/FeTe are plotted in Fig. 3(a) and (b), respectively. For both samples, the phase diagram clearly exhibits the anisotropy between the parallel and perpendicular field for both as-grown and after-two-years cases. Meanwhile, the anisotropy ratio $H_{c2}^{//} / H_{c2}^{\perp}$ of the after-two-years sample shows a decrease, especially for $H_{zero}$, comparing with the as-grown one. For both field directions, the as-grown $H_{max}$ locates between the after-two-years $H_{max}$ and $H_{mid}$, and all three curves show the same variation trend as magnetic field changes. It further manifests that the new feature at 13 K, *i.e.*, the after-two-years $H_{max}$, are relevant to the interfacial superconductivity of the heterostructure. In the study of BiS$_2$-based superconductor LaO$_{0.5}$F$_{0.5}$BiS$_2$, two-step drop of $R(T)$ in different magnetic fields was also observed.[38] Since the LaO$_{0.5}$F$_{0.5}$BiS$_2$ sample was polycrystalline, the origin of two anisotropic upper critical fields was attributed to the anisotropy of the grains in different directions.[38] However, this scenario cannot be applied easily to our heterostructures, since they are all composed of single crystallined films. In addition, the anisotropic behavior of upper critical fields in our heterostructures can be largely suppressed and affected by the annealing process in N$_2$ atmosphere, at the expense of lowering the superconductivity transition temperature (see Supplementary information).

Comparing with type-I superconductors, our Bi$_2$Te$_3$/FeTe heterostructure samples show a relatively broad superconducting transition regime (>3 K) in magnetic fields even for the as-grown samples. This means the mixed states exist in the transition regime and vortex dynamics studies will be important and can provide useful information about the interfacial superconducting behavior. Thermally activated flux flow (TAFF) describes the motion of vortices due to the activation over some energy barriers, *e.g.* pinning centers[39-41]. It was widely studied in 3D iron-based superconductors, such as *β*-FeSe single crystal,[41] $Fe_{1.03}Te_{0.55}Se_{0.45}$,[42] $Fe_{1.14}Te_{0.91}S_{0.09}$,[43] $NdFeAsO_{0.7}F_{0.3}$,[44] and cuprates superconductors, such as, $Bi_2Sr_2CaCu_2O_{8+\delta}$,[45] and $YBa_2Cu_3O_{7-\delta}$,[46]. Recently, TAFF was also reported in novel 2D superconductors, such as FeSe single layer,[47] and exfoliated $NbSe_2$,[48] where in the latter case the TAFF behavior was reported to come from the unbinding of vortex-antivortex pairs. According to the TAFF theory, the resistivity $\rho$ in the TAFF region follows the Arrhenius relation[39-41]

$$\rho(T,H) = \rho_0(H)\exp(-U(H)/T) \qquad (1)$$

where $\rho_0$ is a temperature independent constant, $U$ is the thermal activation energy of the flux flow. Therefore, Equation (1) can be written as $\ln\rho(T,H) = \ln\rho_0(H) - U(H)/T$. At a fixed magnetic field, $\ln\rho(T) - 1/T$ plot is expected to have a linear relation in the TAFF regime. Two samples, $Bi_2Te_3$(7 nm)/FeTe and $Bi_2Te_3$(14 nm)/FeTe, are carefully studied by applying the TAFF theory, as shown in Fig. 4. For both as-grown and after-two-years cases, the temperature dependent resistances of two samples in different perpendicular fields are plotted on $\ln\rho(T,H) - 1/T$ axes in Fig. 4(a), (b), (d) and (e), respectively. As can be seen, all curves exhibit good linear behaviors at low temperature region, manifesting that they follow the Arrhenius relation very well. All fitting lines in different fields cross to one point, whose corresponding temperature $T_m$ should be equal to the $T_c$ of the system. For the as-grown case, $T_m$ of Bi₂Te3(7 nm)/FeTe and $Bi_2Te_3$(14 nm)/FeTe are obtained as 11.6 K and 10.2 K, which are close to their $T_{max}$ = 12.4 K and 11.4 K, respectively. However, for after-two-years results (*c.f.* Fig. 4(b) and (e)), $T_m$ of $Bi_2Te_3$(7 nm)/FeTe and $Bi_2Te_3$(14 nm)/FeTe are obtained as 9 K and 8.7 K, which are closer to their $T_{mid}$ of ~ 11 K than $T_{max}$ of ~ 13 K. This indicates that as-grown samples fall into the TAFF region much faster than the after-two-years samples when the superconducting transition commences. This slower approach to the TAFF region in the after-two-years samples, to a large extent, is affected by the new feature around 13 K, although the origin of this new feature remains unclear.

From the linear fitting of $\ln\rho(T,H)-1/T$ curves, the activation energy $U(H)$ can be obtained from the slope value. Fig. 4(c) and (f) displays $U$ at different magnetic fields for $Bi_2Te_3$(7 nm)/FeTe and $Bi_2Te_3$(14 nm)/FeTe, respectively. For the as-grown samples (triangular symbols in Fig. 4(c) and (f)), $U$ exhibits a logarithmic dependence on the magnetic field, $U=U_0\ln(H_0/H)$, with $H_0 \approx H_{c2}$, as observed in other 2D systems[48-50]. This observation of a 2D vortex system is consistent with the earlier report of the Berezinsky-Kosterlitz-Thouless transition in the $Bi_2Te_3$/FeTe interface.[33,51] For both $Bi_2Te_3$(7 nm)/FeTe and $Bi_2Te_3$(14 nm)/FeTe samples, the fitted value of $H_0$ decreases after two years' air exposure as displayed in Fig. 4(c) and (f), which shows a good agreement with the result of upper critical field $H_{zero}$ in Fig. 3. The energy prefactor $U_0 \sim \Xi \cdot d/\lambda^2$, where $d$ is the superconducting layer thickness, $\lambda$ is the penetration depth and $\Xi$ is a numerical factor which depends on the nature of the energy barrier.[50] In the aged samples, $U$ shows an overall decrease, implying that the aging weakens the vortex pinning behavior or enhances the flux flow of the system. Interestingly, the logarithmic dependence of $U$ on $H$ remains valid, albeit with a lower $U_0$ (circular symbols in Fig. 4(c) and (f)). Therefore, the 2D nature of the vortex system remains robust in the aged heterostructures. Assuming that the dominant thermal activation mechanism remains the same in the aged samples, the drop in $U_0$ compared with the as-grown samples can be attributed to an increase in $\lambda$. Empirically, $\lambda(0) \approx 1.05 \times 10^{-3}(\rho_0/T_c)^{1/2}$, where $\rho_0$ is the residual resistivity of the normal state and $\lambda(0)$ is the zero temperature limit of the penetration depth.[52] Further assuming that the temperature dependences of $\lambda(T)$ and $\rho(T)$ do not vary strongly with age, we can estimate $U_0'/U_0 \sim (\lambda/\lambda')^2 \sim [R(18K)/R'(18K)] \times (T_c'/T_c)$, where the primed quantities are for the aged sample. For Bi$_2$Te$_3$(7 nm)/FeTe, take $(R(18K), R'(18K), T_c, T_c')$ = (21.0 Ω, 25.3 Ω, 10 K, 6 K), $U_0'/U_0$ is estimated to be ~ 0.50, in reasonable agreement with the observed ratio of ~ 0.38. Following the same procedure for Bi$_2$Te$_3$(14 nm)/FeTe, $U_0'/U_0$ is estimated to be ~ 0.52 whereas the observed ratio is also ~ 0.38.

One possible scenario responsible for the logarithmic magnetic field dependence of $U$ is the nucleation and the subsequent motion of the dislocation pairs associated with the vortex lattice. In this model, $U$ is primarily the energy cost of nucleating the pair.[49,50] In the aged samples, the lowering of $U$ can be associated with the relative ease of nucleating the dislocation pairs. Since exposing Bi$_2$Te$_3$ and FeTe to air inevitably promotes the formation of Te vacancies,[35-37] the excess vacancies thus lower the energy barrier required to nucleate dislocation pairs. In addition, these vacancies introduce weak disorder to the material system, thereby resulting in a lower superconducting transition temperature and a higher normal state resistance; these trends are fully consistent with experimental observation in all aged Bi$_2$Te$_3$/FeTe heterostructures.

**Conclusion**

In conclusion, we study the superconducting properties of as-grown and aged Bi$_2$Te$_3$/FeTe heterostructures. Superconductivity is robust after two years' air exposure, even when the thickness of the Bi$_2$Te$_3$ layer is down to 3 nm. Comparing with the upper critical fields of the as-grown measurements, a new feature around 13 K induced by the aging effect is demonstrated to be relevant to the interfacial superconductivity. The resistance of the Bi$_2$Te$_3$/FeTe heterostructures below the superconducting transition temperature obeys the Arrhenius relation, which demonstrates the TAFF behaviour. The activation energy $U(H)$ follows a logarithmic dependence on the applied magnetic field in the as-grown samples, indicating that the vortex system is two-dimensional. The logarithmic dependence remains valid in the aged samples, although $U(H)$ becomes lower at all magnetic fields studied, leading to the conclusion that the 2D vortex system in $Bi_2Te_3$/FeTe heterostructures is robust.

**Methods**

The $Bi_2Te_3$/FeTe heterostructure samples used in the experiment were grown by molecular beam epitaxy on a (111) semi-insulating GaAs substrate with an undoped 50 nm-thick ZnSe buffer layer. A 140 nm-thick FeTe layer was first deposited onto the ZnSe buffer, followed by a growth of the $Bi_2Te_3$ layer on the FeTe layer via van der Waals expitaxy. The thicknesses of the $Bi_2Te_3$ epilayer were 3 nm, 5 nm, 7 nm and 14 nm for four different wafers, respectively. Detailed structural characterizations can be found in the early work.[33] Silver paste and aluminum wires were employed to serve as the electrical contacts, after the wafers were cut into 2 mm × 6 mm strips by a diamond scribe. After the first round measurements on the as-grown heterostructures, all samples were exposed to the air atmosphere at room temperature for two years. To avoid the complication from the sample dependence, all four samples, *i.e.* $Bi_2Te_3$(3 nm)/FeTe, $Bi_2Te_3$(5 nm)/FeTe, $Bi_2Te_3$(7 nm)/FeTe, and $Bi_2Te_3$(14 nm)/FeTe, which were measured in the second round after two years, are exactly the same strips as the ones used in the first round. All transport measurements were conducted in a Quantum Design physical property measurement system with a 14-Tesla superconducting magnet and a base temperature of 2 K.

**References**

1  Hasan, M. Z. & Kane, C. L. Topological insulators. *Rev. Mod. Phys.* **82**, 3045-3067 (2010).

2  Qi, X. L. & Zhang, S. C. Topological insulators and superconductors. *Rev. Mod. Phys.* **83**, 1057-1110 (2011).

3  Zhang, H. J., *et al.* Topological insulators in $Bi_2Se_3$, $Bi_2Te_3$ and $Sb_2Te_3$ with a single Dirac cone on the surface. *Nat. Phys.* **5**, 438-442 (2009).

4  Chen, Y. L., *et al.* Experimental realization of a three-dimensional topological insulator, $Bi_2Te_3$. *Science* **325**, 178-181 (2009).

5  Checkelsky, J. G., *et al.* Quantum interference in macroscopic crystals of nonmetallic $Bi_2Se_3$. *Phys. Rev. Lett.*, **103**, 246601 (2009).

6  Peng, H. L., *et al.* Aharonov–Bohm interference in topological insulator nanoribbons. *Nat. Mater.* **9**, 225 (2010).

7  Gehring, P., *et al.* A natural topological insulator. *Nano Lett.* **13**, 1179–1184 (2013).

8  He, H. T., *et al.* Impurity effect on weak antilocalization in the topological insulator $Bi_2Te_3$. *Phys. Rev. Lett.* **106**, 166805 (2011).

9  Lu, H. Z., Shi, J. & Shen, S. Q. Competition between weak localization and antilocalization in topological surface states. *Phys. Rev. Lett.* **107**, 076801 (2011).

10  Cao, H. L., *et al.* Quantized Hall Effect and Shubnikov–de Haas oscillations in highly doped $Bi_2Se_3$: evidence for layered transport of bulk carriers. *Phys. Rev. Lett.* **108**, 216803 (2012).

11  Cha, J. J., *et al.* Weak antilocalization in $Bi_2(Se_xTe_{1-x})_3$ nanoribbons and nanoplates. *Nano Lett.*, **12**, 1107-1111 (2012).

12  Lang, M. R., *et al.* Competing weak localization and weak antilocalization in ultrathin topological insulators. *Nano Lett.* **13**, 48-53 (2013).

13  Liu, H. C., *et al.* Tunable interaction-induced localization of surface electrons in antidot nanostructured $Bi_2Te_3$ thin films. *ACS Nano* **8**, 9616-9621 (2014).

14  Fu, L. & Kane, C. L. Superconducting proximity effect and Majorana Fermions at the surface of a topological insulator. *Phys. Rev. Lett.* **100**, 096407 (2008).

15  Hor, Y. S., *et al.* Superconductivity in $Cu_xBi_2Se_3$ and its implications for pairing in the undoped topological insulator. *Phys. Rev. Lett.* **104**, 057001 (2010).

16  Kriener, M., *et al.* Bulk superconducting phase with a full energy gap in the doped topological insulator $Cu_xBi_2Se_3$. *Phys. Rev. Lett.* **106**, 127004 (2011).

17  Shen, J., *et al.* Revealing surface states in In-doped SnTe nanoplates with low bulk mobility. *Nano Lett.* **15**, 3827-3832 (2015).

18  Liu, Z. K., *et al.* A stable three-dimensional topological Dirac semimetal $Cd_3As_2$. *Nat. Mater.* **13**, 677-681 (2014).

19  Xu, S. Y., *et al.* Observation of Fermi arc surface states in a topological metal. *Science*, **347**, 294-298 (2015).

20  Weng, H. M., *et al.* Weyl semimetal phase in noncentrosymmetric transition-metal monophosphides. *Phys. Rev. X* **5**, 011029 (2015).

21  Xiong, J., *et al.* Evidence for the chiral anomaly in the Dirac semimetal $Na_3Bi$. *Science* **350**, 413-416 (2015).

22  Li, H., *et al.* Negative magnetoresistance in Dirac semimetal $Cd_3As_2$. *Nat. Commun.* **6**, 10301 (2016).

23  Zhang, D. M., *et al.* Superconducting proximity effect and possible evidence for Pearl vortices in a candidate topological insulator. *Phys. Rev. B*, **84**, 165120 (2011).

24  Sochnikov, I., *et al.* Direct measurement of current-phase relations in superconductor/topological insulator/superconductor junctions. *Nano Lett.* **13**, 3086-3092 (2013).

25  Zareapour, P., *et al.* Proximity-induced high-temperature superconductivity in the topological insulators $Bi_2Se_3$ and $Bi_2Te_3$. *Nat. Commun.* **3**, 1056, (2012).

26  Williams, J. R., *et al.* Unconventional Josephson Effect in hybrid superconductor-topological insulator devices. *Phys. Rev. Lett.* **109**, 056803 (2012).

27  Veldhorst, M., *et al.* Josephson supercurrent through a topological insulator surface state. *Nat. Mat.* **11**, 417-421 (2012).

28  Wang, M. X., *et al.* The coexistence of superconductivity and topological order in the $Bi_2Se_3$ thin films. *Science* **336**, 52-55 (2012).

29  Xu, J. P., *et al.* Experimental detection of a Majorana mode in the core of a magnetic vortex inside a topological insulator-superconductor $Bi_2Te_3$/$NbSe_2$ heterostructure. *Phys. Rev. Lett.* **114**, 017001 (2015).

30  Mizukami, Y., *et al.* Extremely strong-coupling superconductivity in artificial two-dimensional Kondo lattices. *Nat. Phys.* **7**, 849 (2011).

31  Goh, S. K., *et al.* Anomalous upper critical field in $CeCoIn_5$/$YbCoIn_5$ superlattices with a Rashba-type heavy fermion interface. *Phys. Rev. Lett.* **109**, 157006 (2012).

32  Shimozawa, M., *et al.* Controllable Rashba spin-orbit interaction in artificially engineered superlattices involving the heavy-fermion superconductor $CeCoIn_5$. *Phys. Rev. Lett.* **112**, 156404 (2014).

33  He, Q. L., *et al.* Two-dimensional superconductivity at the interface of a $Bi_2Te_3$/FeTe heterostructure. *Nat. Commun.* **5**, 4247 (2014).

34  Beenakker, C. W. J. Search for Majorana Fermions in superconductors. *Annu. Rev. Condens. Matter Phys.* **4**, 113 (2013).

35  Bando, H., *et al.* The time-dependent process of oxidation of the surface of $Bi_2Te_3$ studied by x-ray photoelectron spectroscopy. *J. Phys.: Condens. Matter* **12**, 5607-5616 (2000).

36  Zhao, Y. X. & Burda, C. Chemical Synthesis of $Bi_{0.5}Sb_{1.5}Te_3$ nanocrystals and their surface oxidation properties. *ACS Appl. Mat. & Interfaces* **1**, 1259-1263 (2009).

37  Haindl, S., *et al.* Thin film growth of Fe-based superconductors: from fundamental properties to functional devices. A comparative review. *Rep. Prog. Phys.* **77**, 046502 (2014).

38  Mizuguchi, Y., *et al.* Anisotropic upper critical field of the BiS$_2$-based superconductor LaO$_{0.5}$F$_{0.5}$BiS$_2$. *Phys. Rev. B* **89**, 174515 (2014).

39  Palstra, T. T. M., Thermally activated dissipation in Bi$_{2.2}$Sr$_2$Ca$_{0.8}$Cu$_2$O$_{8+\delta}$. *Phys. Rev. Lett.* **61**, 1662 (1988).

40  Blatter, G., *et al.* Vortices in high-temperature superconductors. *Rev. Mod. Phys.* **66**, 1125 (1994).

41  Lei, H. C., *et al.* Critical fields, thermally activated transport, and critical current density of *β*-FeSe single crystals. *Phys. Rev. B* **84**, 014520 (2011).

42  Ge, Y. J., *et al.* Superconducting properties of highly oriented Fe$_{1.03}$Te$_{0.55}$Se$_{0.45}$ with excess Fe. *Solid State Commun.* **150**, 1641-1645 (2010).

43  Lei, H. C., *et al.* Thermally activated energy and flux-flow Hall effect of Fe$_{1+y}$(Te$_{1+x}$S$_x$)$_z$. *Phys. Rev. B* **82**, 134525 (2010).

44  Jaroszynski, J., *et al.* Upper critical fields and thermally-activated transport of NdFeAsO$_{0.7}$F$_{0.3}$ single crystal. *Phys. Rev. B* **78**, 174523 (2008).

45  Zhang, Y. Z., *et al.* Deviations from plastic barriers in Bi$_2$Sr$_2$CaCu$_2$O$_{8+\delta}$ thin films. *Phys. Rev. B* **71**, 052502 (2005).

46  Zhang, Y. Z., *et al.* Thermally activated energies of YBa$_2$Cu$_3$O$_{7-\delta}$ and Y$_{0.8}$Ca$_{0.2}$Ba$_2$Cu$_3$O$_{7-\delta}$ thin films. *Phys. Rev. B*, **74**, 144521 (2006).

47  Sun, Y., *et al.* High temperature superconducting FeSe films on SrTiO$_3$ substrates. *Sci. Rep.* **4**, 6040 (2014).

48  Tsen, A. W., *et al.* Nature of the quantum metal in a two-dimensional crystalline superconductor. *Nat. Phys.* **12**, 208-212, (2015).

49  Feigel'man, M.V., *et al.* Pinning and creep in layered superconductors. *Physica C* **167**, 177-187 (1990).

50  Qiu, X. G., *et al.* Two different thermally activated flux-flow regimes in oxygen-deficient

YBa$_2$Cu$_3$O$_{7-x}$ thin films. *Phys. Rev. B* **52**, 559 (1995).

51  Kunchur, M. N., *et al.* Current-induced depairing in the Bi$_2$Te$_3$/FeTe interfacial superconductor. *Phys. Rev. B* **92**, 094502 (2015).

52  Kes, P. H. & Tsuei, C. C. Two-dimensional collective flux pinning, defects, and structural relaxation in amorphous superconducting films. *Phys. Rev. B* **28**, 5126 (1983).

**Acknowledgements**

This work was supported by CUHK (Startup Grant, Direct Grant No. 4053123), UGC Hong Kong (ECS/24300214), and the Research Grants Council of the Hong Kong under Grant Nos. 16305514, 17303714, 16304515 and AoE/P-04/08.

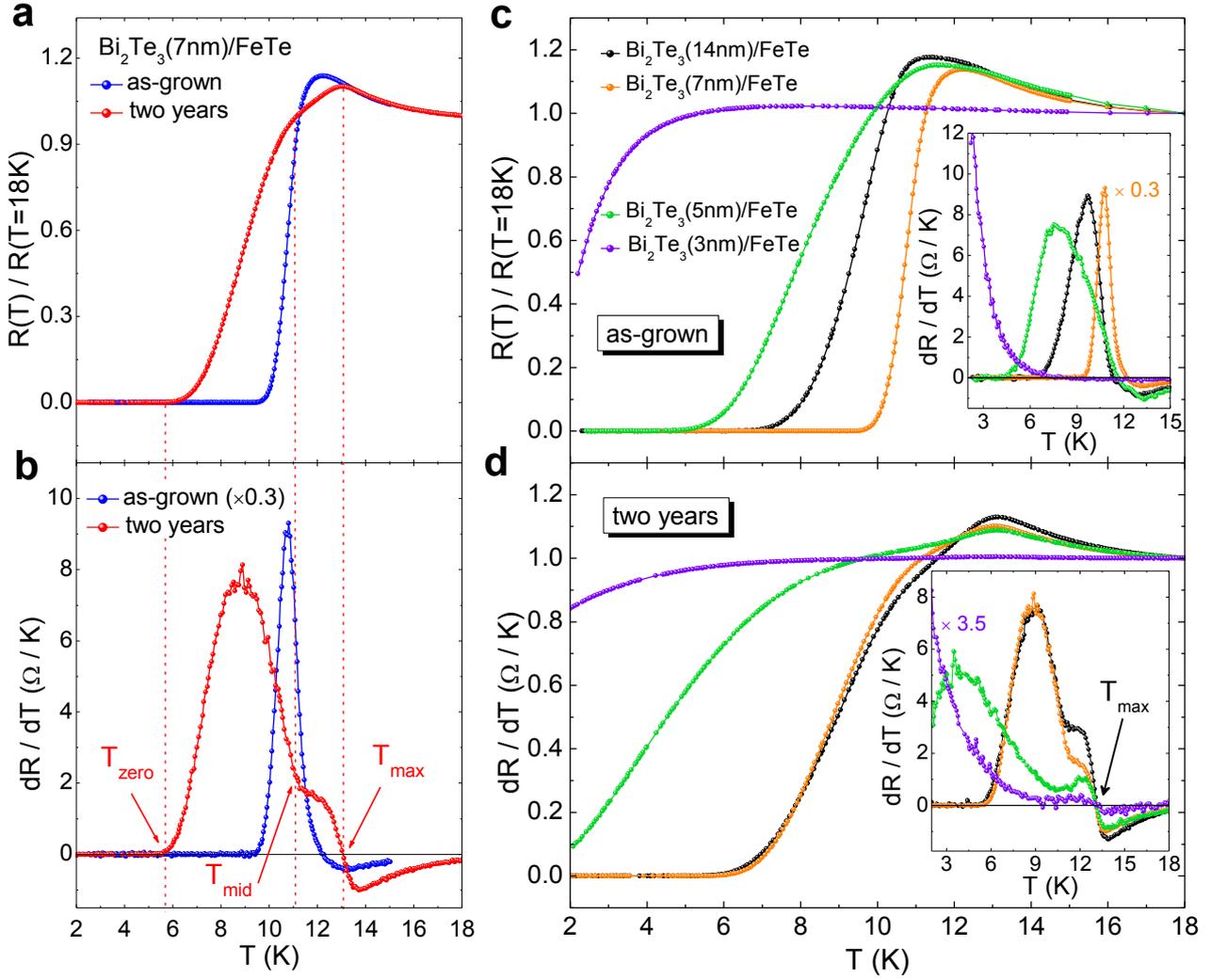

Figure 1. (a) Normalized temperature dependent resistances and (b) *dR*/*dT* curves of sample Bi$_2$Te$_3$(7 nm)/FeTe in as-grown and after-two-years cases. The as-grown and after-two-years normal state resistances at 18 K are 21.0 Ω and 25.3 Ω, respectively. The $T_{max}$, $T_{mid}$, $T_{zero}$ of after-two-years case are indicated with dash lines and arrows. Normalized temperature dependent resistances of samples Bi$_2$Te$_3$(14 nm)/FeTe, Bi$_2$Te$_3$(7 nm)/FeTe, Bi$_2$Te$_3$(5 nm)/FeTe and Bi$_2$Te$_3$(3 nm)/FeTe in (c) as-grown and (d) after-two-years case, respectively. The corresponding *dR*/*dT* curves are given in their insets.

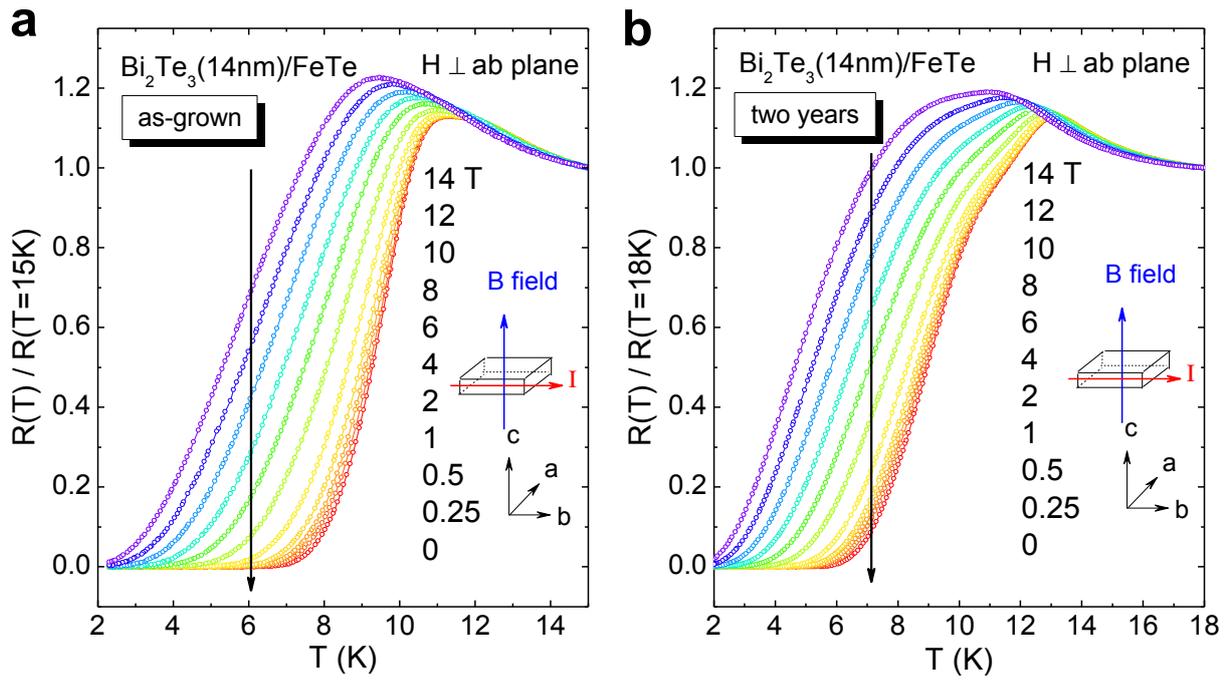

Figure 2. Normalized temperature dependent resistances of sample Bi$_2$Te$_3$(14 nm)/FeTe in magnetic fields ranging from 0 T to 14 T in (a) as-grown and (b) after-two-years cases, respectively. The magnetic field is perpendicular to the *ab* plane as shown in the insets.

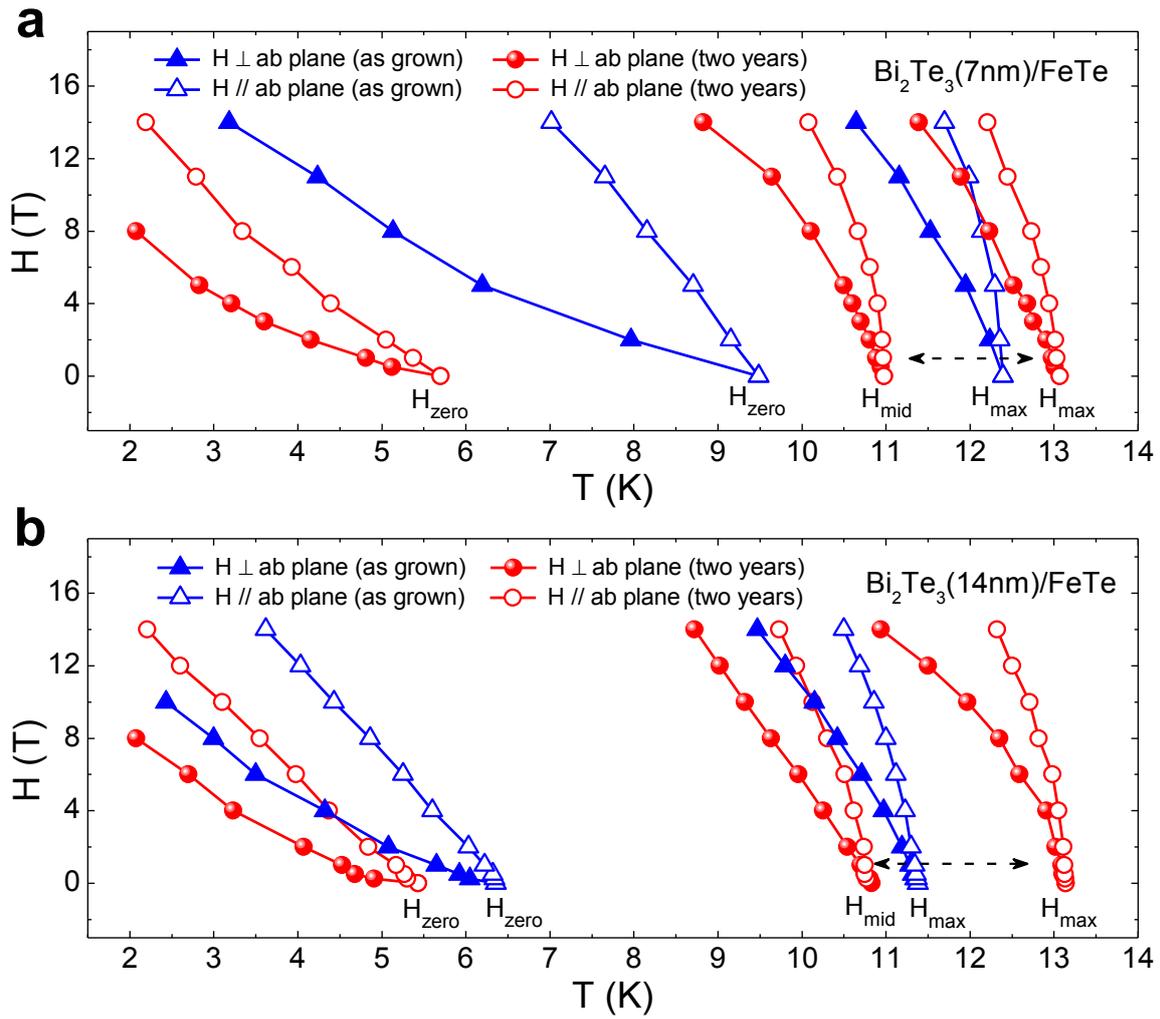

Figure 3. Magnetic field-temperature phase diagram of samples (a) $Bi_2Te_3$(7 nm)/FeTe and (b) $Bi_2Te_3$(14 nm)/FeTe. The upper critical field $H_{max}$ and $H_{zero}$ in the as-grown case are plotted as triangle symbols, and the $H_{max}$, $H_{mid}$ and $H_{zero}$ in the after-two-years case are presented as circle symbols. The solid and hollow symbols represent the perpendicular and parallel fields' situations, respectively.

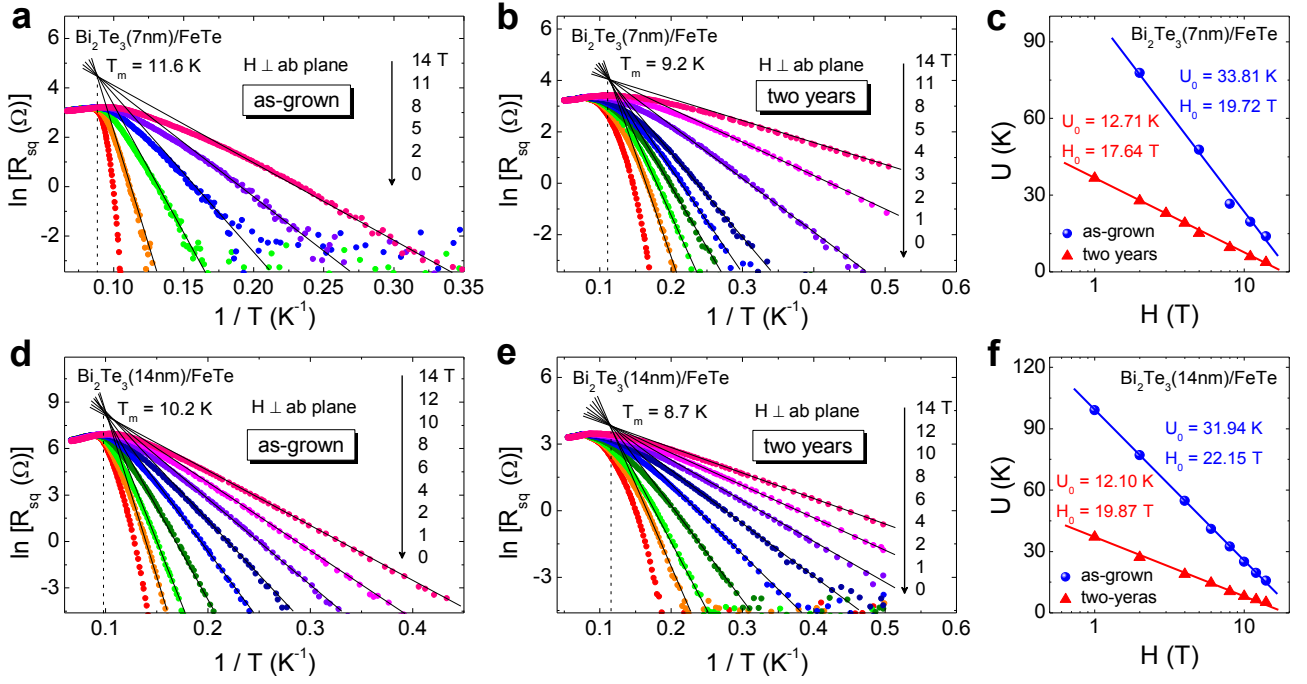

Figure 4. ln$R_{sq}(T)$ vs. 1/$T$ in different perpendicular magnetic fields for sample (a) Bi$_2$Te$_3$(7 nm)/FeTe in as-grown case, (b) Bi$_2$Te$_3$(7 nm)/FeTe in after-two-years case, (d) Bi$_2$Te$_3$(14 nm)/FeTe in as-grown case, (e) Bi$_2$Te$_3$(14 nm)/FeTe in after-two-years case. The solid lines in (a), (b), (d), (e) are fitting results from the Arrhenius relation, whose slopes give the values of $U$ in (c) and (f). The solid lines in (c) and (f) are fitting results from the function $U = U_0 \ln(H_0/H)$.

# Supplementary information for "Robust two-dimensional superconductivity and vortex system in $Bi_2Te_3$/FeTe heterostructures"

Hong-Chao Liu[1,2], Hui Li[1], Qing Lin He[3], Iam Keong Sou[1,3], Swee K. Goh[2,*], and Jiannong Wang[1,3,*]

[1]*Department of Physics, The Hong Kong University of Science and Technology, Clear Water Bay, Hong Kong, China*

[2]*Department of Physics, The Chinese University of Hong Kong, Shatin, New Territories, Hong Kong, China*

[3]*William Mong Institute of Nano Science and Technology, The Hong Kong University of Science and Technology, Clear Water Bay, Hong Kong, China*

*corresponding author: skgoh@phy.cuhk.edu.hk, phjwang@ust.hk

The annealing effect on $Bi_2Te_3$(7 nm)/FeTe samples are studied. Here, $Bi_2Te_3$(7 nm)/FeTe #S1 and #S2 samples come from the same wafer as the one in the manuscript after two years' air exposure but without as-grown measurements. Figure S1(a) shows the $R(T)$ curves of sample #S1 annealing in the $N_2$ atmosphere at 100 °C with different time. The corresponding $dR/dT$ curves are plotted in Figure S1(b). From Figure S1(b), we can see that a shoulder-like structure appears around 11 K – 13.5 K in $dR/dT$ curve without the annealing process, which is consistent with the result in the manuscript. By increasing the anneal time and holding the temperature at 100 °C, this shoulder-like structure weakens gradually. Furthermore, the shoulder-like structure is completely suppressed after 160 mins annealing holding at 100 °C. Meanwhile, as the annealing time increases, the transition temperature continues to decrease and a new $T_{mid}$ with an unknown origin appears around 9 K, as shown in Figure S1(a) and (b). Moreover, the heterostructure seems to reach a stable state after 160 mins anneal at 100 °C, because a further time increase from 160 to 1060 mins gives nearly overlapped $R(T)$ curves.

Although the heterostructure can still reach a completely superconducting state above 4 K after 10 hours annealing at 100 °C, a higher annealing temperature will strongly suppressed its superconductivity. As shown in Figure S1(c), after the anneal at 160 °C around 2.5 hours, the transition temperature of sample #S2 decreases below 7 K and cannot fully develop into the superconducting state above 2 K. Our annealing effect study suggests that the micro/nano device fabrication, for example, the photolithography or electron beam lithography, which cannot avoid the high temperature processes, may do harm to the superconductivity of the $Bi_2Te_3$/FeTe heterostructure. This also gives an explanation that why we used diamond scribe to prepare the samples and used silver paste to achieve the wire connections in our experiments.

The annealing effect on the flux flow behavior is also studied on $Bi_2Te_3$(7 nm)/FeTe #S1, as shown in Fig. S2. For both before and after annealing situations, the temperature dependent resistance follows the Arrhenius relation (*c.f.* Fig. S2 (a) and (b)), and the activation energy $U(H)$ follows the logarithmic dependence on the applied magnetic field (*c.f.* Fig. S2 (c)). This indicates that the 2D vortex system is still robust after the annealing process although the value of $U_0$ decreases and the flux flow behavior of the system is enhanced.

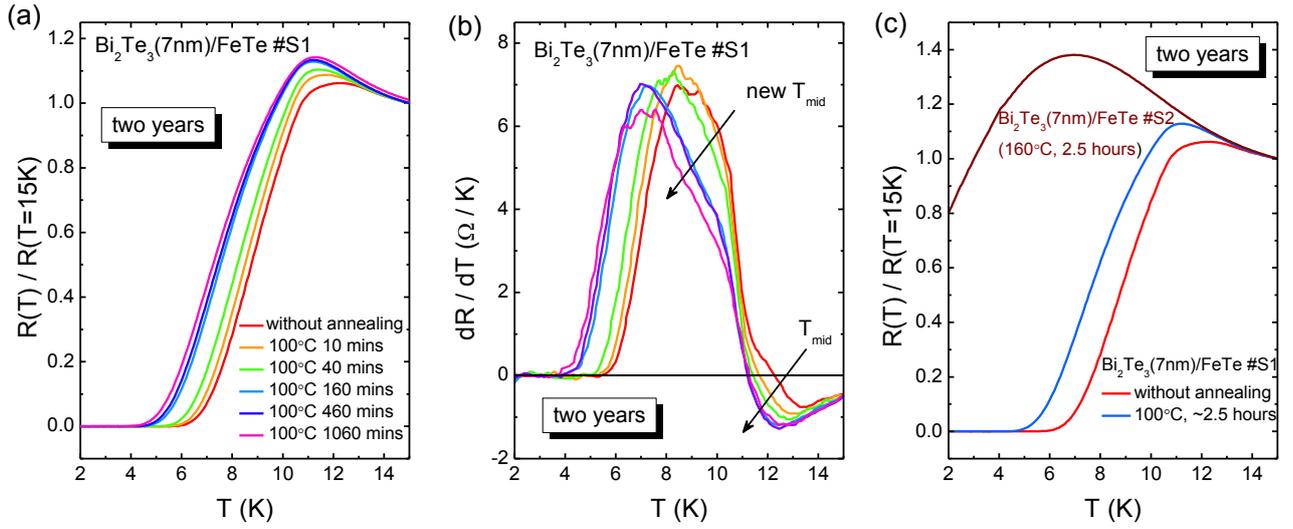

Figure S1. Annealing effect on after-two-years sample $Bi_2Te_3$(7 nm)/FeTe #S1 and #S2. (a) Sample $Bi_2Te_3$(7 nm)/FeTe #S1 annealed in the $N_2$ atmosphere at 100 $^o$C with different time. (b) $dR/dT$ curves corresponding to the $R(T)$ curves in (a). (c) Sample $Bi_2Te_3$(7 nm)/FeTe #S1 and #S2 annealed in the $N_2$ atmosphere at 100 $^o$C 2.5 hours, and 160 $^o$C 2.5 hours, respectively.

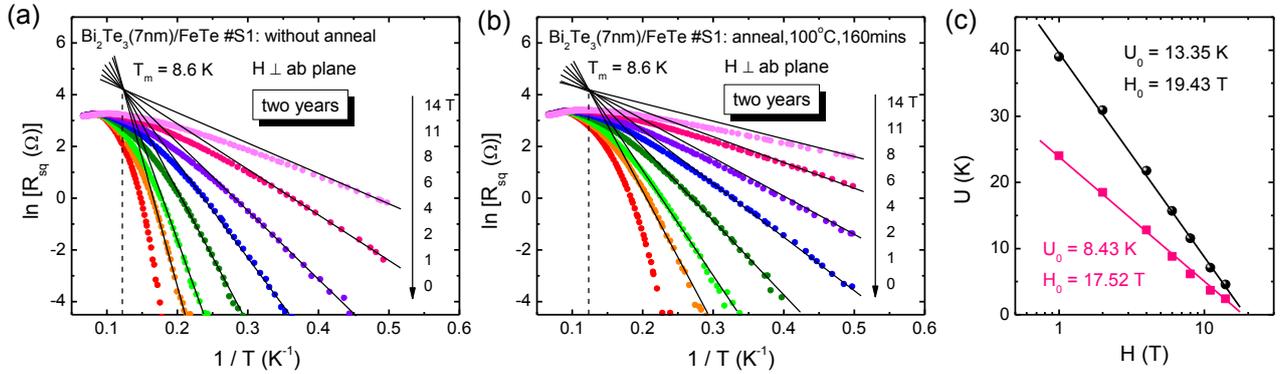

Figure S2. $\ln R_{sq}(T)$ vs. $1/T$ in different perpendicular magnetic fields for sample $Bi_2Te_3$(7 nm)/FeTe #S1 (a) without anneal; (b) anneal at 100$^o$C, 160 mins. The solid lines in (a), (b) are fitting results from the Arrhenius relation, whose slopes give the values of $U$ in (c). The solid lines in (c) are fitting results from the function $U = U_0 \ln(H_0/H)$.